\documentclass[twocolumn]{aastex61}
\usepackage{graphicx}
\usepackage{amsmath}
\usepackage{rotating}

\def\stacksymbols #1#2#3#4{\def\theguybelow{#2}
        \def\verticalposition{\lower#3pt}
        \def\spacingwithinsymbol{\baselineskip0pt\lineskip#4pt}
        \mathrel{\mathpalette\intermediary#1}}
\def\intermediary #1#2{\verticalposition\vbox{\spacingwithinsymbol
        \everycr={}\tabskip0pt
        \halign{$\mathsurround0pt#1\hfil##\hfil$\crcr#2\crcr
                \theguybelow\crcr}}}

\shorttitle{The Cooling Flow Problem in the Milky Way}
\shortauthors{Fang et al.}

\begin{document}

\title {On the Cooling Flow Problem in the Gaseous Halo of the Milky Way}

\correspondingauthor{Fulai Guo}
\email{fulai@shao.ac.cn}

\author{Xiang-Er Fang}
\affil{Key Laboratory for Research in Galaxies and Cosmology, Department of Astronomy, University of Science and Technology of China, Hefei, Anhui 230026, China}

\author{Fulai Guo}
\affiliation{Key Laboratory for Research in Galaxies and Cosmology, Shanghai Astronomical Observatory, Chinese Academy of Sciences, 80 Nandan Road, Shanghai 200030, China}
\affiliation{University of Chinese Academy of Sciences, 19A Yuquan Road, 100049, Beijing, China}

\author{Ye-Fei Yuan}
\affil{Key Laboratory for Research in Galaxies and Cosmology, Department of Astronomy, University of Science and Technology of China, Hefei, Anhui 230026, China}

 \begin{abstract}
Theoretical and observational arguments suggest that there is a large amount of hot ($\sim 10^6$ K), diffuse gas residing in the Milky Way's halo, while its total mass and spatial distribution are still unclear. In this work, we present a general model for the gas density distribution in the Galactic halo, and investigate the gas evolution under radiative cooling with a series of 2D hydrodynamic simulations. We find that the mass inflow rate in the developed cooling flow increases with gas metallicity and the total gas mass in the halo. For a fixed halo gas mass, the spatial gas distribution affects the onset time of the cooling catastrophe, which starts earlier when the gas distribution is more centrally-peaked, but does not substantially affect the final mass inflow rate. The gravity from the Galactic bulge and disk affects gas properties in inner regions, but has little effect on the final inflow rate either. We confirm our results by investigating cooling flows in several density models adopted from the literature, including the Navarro-Frenk-White (NFW) model, the cored-NFW model, the Maller \& Bullock model, and the $\beta$ model. Typical mass inflow rates in our simulations range from $\sim 5 M_{\odot}$ yr$^{-1}$ to $\sim 60 M_{\odot}$ yr$^{-1}$, and are much higher than the observed star formation rate in our Galaxy, suggesting that stellar and active galactic nucleus feedback processes may play important roles in the evolution of the Milky Way (MW) and MW-type galaxies.
\end{abstract}


\section{Introduction}
\label{section1}
Several independent observational studies, such as X-ray observations, ram-pressure stripping of Milky Way satellite galaxies, pulsar dispersion measure measurements, indicate that there exists a significant reservoir of hot baryons in the Milky Way (MW) halo (e.g., \citealt{MillerBregman2015}; \citealt{GrcevichPutman2009}; \citealt{Gatto2013}; \citealt{Salem2015}; \citealt{Fang2013}). However, the total mass of the hot gaseous halo and its spatial temperature and density distributions are still very difficult to determine to date. The structure and evolution of the hot gaseous halo may play a very important role in the formation and evolution of the MW. 

The total mass of the hot halo gas may account for a significant fraction or even all of the ``missing baryons" of our Galaxy \citep{Fang2013}. To date various observational evidences including maser observations, satellite kinematics constrain the MW's virial mass in the range of $M_v=(1.0-2.4)\times 10^{12}M_\odot$ (\citealt{Boylan-Kolchin2013}). In the absence of mass loss and taking the cosmological baryon fraction $f_b=0.157$ recently measured by {\it Planck} (\citealt{planck16}), the MW's baryonic mass is $M_b=f_b M_v \simeq (1.57-3.77)\times 10^{11}M_\odot$. However, in the MW, the observed cold baryonic mass is $M_{\rm cold} \simeq 0.65\times 10^{11}M_\odot$(\citealt{McMillan2012}), suggesting that $\gtrsim 10^{11}M_\odot$ of baryons are missing within our MW's virial radius. A large fraction of these missing baryons may reside in the MW's extended halo in the form of hot gas \citep{Fang2013}.

The properties of the hot gaseous halo have been investigated mainly through two approaches: ram-pressure stripping of dwarf satellite galaxies in the MW halo and X-ray spectroscopic observations. The former method is circumstantial, based on the assumption that the lack of HI gas in MW dwarfs is due to ram pressure stripping by the halo gas. Using this method, \citealt{GrcevichPutman2009} find that the halo gas density is greater than $2 \times 10^{-4}$ cm$^{-3}$ out to galactocentric distances of at least $70$ kpc. This result is further confirmed by a more recent constraint of the halo gas density $n=1.3-3.6\times 10^{-4}$ cm$^{-3}$ at $r=50-90$ kpc when using the same ram-pressure stripping argument to two MW dwarfs: Sextans and Carina (\citealt{Gatto2013}). \citet{Salem2015} use both analytic methods and three-dimensional hydrodynamic simulations to investigate ram-pressure stripping signatures of the Large Magellanic Cloud's gaseous disk, finding that the MW's halo gas density is $n=1.1\begin{subarray}
\ +.44\\-.45
\end{subarray} \times 10^{-4}$ cm$^{-3}$ at $r=48.2\pm 5$ kpc. 

The later method with X-ray spectroscopic observations is direct, but its results rely significantly on several assumptions, including metallicity and the spatial density distribution of the halo gas. In addition to the halo gas, additional components could also contribute to the observed soft X-ray background (SXRB). It is known that the SXRB (between 0.5-2 keV) is composed of three components: (1) the cosmic X-ray background (CXB), which is thought to come mostly from active galactic nuclei (AGNs); (2) local thermal plasma emissions from a combination of the local hot bubble (LHB) and solar wind charge exchange (SWCX) processes. The LHB is believed to be a supernova remnant within which the gas is at around $10^6$ K (\citealt{Snowden1990,Snowden1993} and \citealt{Smith2007}), while the SWCX is the emission from charge exchange reactions between neutral hydrogen and helium atoms and solar wind ions around our solar system (\citealt{Cravens2001}, \citealt{Snowden2004}, \citealt{Koutroumpa2006,Koutroumpa2011}); (3) ``non-local'' thermal plasma emissions from both supernova-driven outflows from the Galactic disk (\citealt{JoungMac2006}, \citealt{Hill2012}) and the more extended gaseous halo. So investigating the hot halo gas through X-ray observations depends on how accurately we can determine and then subtract these additional components. \citet{MillerBregman2013,MillerBregman2015} adopted a $\beta$-model (see Section \ref{section:othermodels}) for the radial density profile of the halo gas and a constant-density LHB to fit a large sample of O VII and O VIII absorption and emission line measurements from {\it XMM-Newton}/EPIC-MOS spectra, and obtained a best-fit $\beta$-model with $\beta =0.50$.

Another direct probe for the halo gas density is the dispersion measure (DM) of pulsars. \citet{Nugaev2015} recently investigated the contribution of the hot gas in the MW's halo to the DM of the pulsars with three models for the halo gas density distribution including the Navarro, Frenk \&White (NFW), Maller \& Bullock (MB), and Feldmann, Hooper \& Gnedin (FHG) profiles. They concluded that the MB and FHG models are compatible with the observed DMs, while the DM of the NFW model is too high to be consistent with observational data. \citet{Fang2013} also obtained a similar result though they additionally considered the contribution of the warm ionized medium (WIM) in and near the Galactic disk to the DM (see Figure 3 in their paper).

The existence of extended gaseous haloes around Milky-way-like galaxies has also been predicted in numerical simulations. For example, \citet{Nuza2014} performed a constrained cosmological simulation of the Local Group (LG) to investigate the gas distribution in the MW halo in three phases: cold, hot and HI. They found that the hot gas dominates in the halo with a total mass of  $M_{\rm hot}\sim 4-5\times 10^{10} M_{\odot}$ and the spatial density distribution is consistent with the $\beta$-model for $r \gtrsim 50$ kpc (see \citealt{MillerBregman2015}).  

The hot diffuse gas has been extensively studied in galaxy clusters with X-ray observations and numerical simulations, and is usually called the intracluster medium (ICM) there (see \citealt{werner19} for a recent review). In cool-core galaxy clusters, the central gas cooling time is very short ($\sim 10^{8}-10^{9}$ yr; e.g., \citealt{Hudson2010}; \citealt{McDonald2017}), and in the absence of heating sources, a cooling catastrophe is expected to develop in central regions of galaxy clusters, resulting in significant cooling flows with typical mass inflow rates of $\sim 100-1000$ $M_{\odot} {\rm ~yr}^{-1}$ (\citealt{White1997}; \citealt{Allen2001}; \citealt{Hudson2010}). Strong cooling flows are inconsistent with multi-wavelength observations of most galaxy clusters \citep{Peterson2006}, and AGN feedback is often invoked to heat the ICM and solve this so-called cooling flow problem (e.g., \citealt{guo08}; \citealt{mcnamara12}; \citealt{guo18}). 
 On the galaxy scale, the development of cooling flows is shown to be bimodal, either resulting in a central cooling catastrophe or remaining in the hot mode with a central cuspy temperature profile (\citealt{Guo2014}; \citealt{guo14b}; \citealt{stern19}).
 If the MW indeed has a significant hot gaseous halo as observations have indicated so far, it would be important to investigate the impact of radiative cooling on the evolution of the halo gas, which may supply cool gas to the Galactic disk in the form of cooling flows. 

In this work, we present a new general model for the radial density distribution of the hot halo gas. Starting from hydrostatic equilibrium in the MW's potential well, we investigate the thermodynamic evolution of the hot gas under radiative cooling using a suite of two-dimensional (2D) hydrodynamic simulations. We pay particular attention to the gas cooling times in different initial gas density profiles and the resulting mass inflow rates toward the Galactic center. We investigate the roles of gas metallicity, the total gas mass in the halo, the spatial distribution of the halo gas, the gravity from the Galactic bulge and disk on the developed cooling flow. We also confirm our main results with several gas density models in the literature. The rest of the paper is organized as follows. In Section \ref{section:method}, we describe our methods in detail, including the basic equations, the gravitational potential of the MW, our model for the density profile of the hot halo gas, simulation setup, and initial conditions. Our results are presented in Section \ref{section:results}. In Section \ref{section:othermodels}, we compare our model with several well-known models in the literature. We summarize our results in Section \ref{section:summary}.

\section{Methods}
\label{section:method}
\subsection{Basic Equations}
\label{equations}

We investigate the evolution of the hot gas in the MW's halo by solving the following hydrodynamic equations:
\begin{eqnarray}
\frac{d \rho}{d t} + \rho \nabla \cdot {\bf v} = 0,\label{hydro1}
\end{eqnarray}
\begin{eqnarray}
\rho \frac{d {\bf v}}{d t} = -\nabla P-\rho \nabla \Phi_{\rm MW} ,\label{hydro2}
\end{eqnarray}
\begin{eqnarray}
\frac{\partial e}{\partial t} +\nabla \cdot(e{\bf v})=-P\nabla \cdot {\bf v}-\mathcal{C},\label{hydro3}
\end{eqnarray}
where $\rho$, ${\bf v}$ and $e$ are the density, velocity and internal energy density of the gas, respectively. $P=(\gamma -1)e$ is the gas thermal pressure, where $\gamma = 5/3$ is the adiabatic index. $\Phi_{MW}$ is the gravitational potential of the MW (see Section~\ref{Gravitation}). $\mathcal{C}=n_{\rm i}n_{\rm e}\Lambda(T, Z)$ is the radiative cooling rate per volume, where $n_{\rm i}$ is the ion number density, $n_{\rm e}$ is the electron number density, and the cooling function $\Lambda(T, Z)$ depending on the gas temperature $T$ and metallicity $Z$ is adopted from \citet{SutherlandDopita1993}. The gas temperature $T$ can be derived from the ideal gas equation of state (EOS):
\begin{equation}
 \label{eos}
 P=\frac{\rho k_{B} T}{\mu m_{p}}{,}
\end{equation}
where $k_{B}$ is Boltzmann's constant, $\mu = 0.59 $ is the mean molecular weight per particle and $m_{p}$ is the proton mass (\citealt{MillerBregman2015}; \citealt{guo08a}).

\subsection{Gravitational Potential of the Milky Way}
\label{Gravitation}

The gravitational potential of the MW is mainly contributed by three components: a dark matter halo, a stellar disk and a spherical stellar bulge:
\begin{equation}
\Phi_{\rm MW} = \Phi_{\rm dm} + \Phi_{\rm disk} + \Phi_{\rm bulge} {.}
\end{equation}
Here we ignore the self-gravity of the hot gas, which is negligible compared to the other components. For simplicity, we assume that $\Phi_{\rm MW}$ is fixed in our simulations.
In outer regions of the Galaxy, the gravitational potential is dominated by dark matter, whose radial density distribution may be described by the NFW profile (\citealt{Navarro1996,Navarro1997}):
\begin{equation}
 \label{rho-dm}
 \rho_{\rm dm}(r) = \frac{m_0/2\pi}{r(r+r_s)^2}{,}
\end{equation}
where $r_s$ is the scale radius of the NFW profile and $m_0$ is a characteristic mass. 

The NFW profile contains two parameters: $r_s$ and $m_0$, and here we instead use the virial mass $M_{\rm vir}$ and the concentration $C_{\rm v}$ to describe it. $M_{\rm vir}$ is the total dark matter mass enclosed in the virial radius $r_{\rm vir}$, within which the mean dark matter density is $\Delta_{\rm vir}$ times the critical density of the Universe $\rho_c=3H(z)^2/8\pi G$. Here $H(z)$ is the Hubble constant at redshift $z$ of the system, $G$ is the gravitational constant, and we choose $\Delta_{\rm vir}=200$. In all our simulations, we take the virial mass to be $M_{\rm vir}=10^{12} M_\odot$.  The virial radius can then be derived:
\begin{equation}
r_{\rm vir} = (\frac{3M_{\rm vir}}{4\pi\Delta_{\rm vir}\rho_c})^{1/3}=206 {\rm ~kpc}
\end{equation}
which is the same as in \citet{DierickxLoeb2017}. The value of the concentration $C_v$ is derived from the correlation between the concentration and the virial mass found in the cosmological simulations of \citet{Duffy2008}:
\begin{equation}
C_v=A(M_{\rm vir}/M_{\rm pivot})^B(1+z)^C
\end{equation}
where $A=5.74$, $B=-0.097$, $z=0$ and $C=0$. The derived concentration is $C_v=6.355$, and thus the scale radius $r_{\rm s}=r_{\rm vir}/C_v=32.5$ kpc.
The  characteristic mass $m_0$ is related to the concentration $C_v$ and the scale radius $r_{\rm s}$ through
\begin{equation}
m_0=2\pi\rho_c\delta_c r_s^3 {,}
\end{equation}
where the characteristic density $\delta_c$ is 
\begin{equation}
\delta_c=\frac{\Delta_{vir}}{3}\frac{C_v^3}{\ln(1+C_v)-C_v/(1+C_v)} {.} 
\end{equation}

Combined with the above equations, the corresponding gravitational potential contributed by dark matter can be written as \citep{Guo2014} :
\begin{equation}
\Phi_{\rm dm}(r)=-\frac{2Gm_0}{r_s}\frac{\ln(1+x)}{x}
\end{equation}
where $x=r/r_{\rm s}$. Unless noted otherwise, we only take into account the contribution of dark matter to the MW's gravitational potential $\Phi_{\rm MW}$ in our simulations, while the impact of the Galactic disk and bulge on our results is specifically investigated in Section \ref{diskbulge}.

The Galactic disk and bulge mainly affect the gravitational potential of inner regions of the MW. For $\Phi_{\rm disk}$ and $\Phi_{\rm bulge}$, we adopt the models used in \citet{HelmiWhite2001} and \citet{guo12}:
\begin{gather}
\Phi_{\rm disk}=-\frac{GM_{\rm disk}}{\sqrt{R^2+(a+\sqrt{z^2+b^2})^2}} {,}\\
\Phi_{\rm bulge}=-\frac{GM_{\rm bulge}}{r+c} {.}
\end{gather}
where $a=6.5$ kpc, $b=0.26$ kpc, $c=0.7$ kpc, $R$ is the galactocentric radius in the Galactic plane, and $z$ is the height from the Galactic disk. For the masses of the disk and bulge, we use the updated values in \citet{McMillan2017}: $M_{\rm bulge}=9.23\times 10^{9}M_\odot$. The MW's stellar disk can be decomposed into two components: the thin and thick disks (e.g. \citealt{GilmoreReid1983}), and the resulting disk mass is $M_{\rm disk}=M_{\rm d,thin}+M_{\rm d,thick}=4.57\times 10^{10}M_\odot$, where $M_{\rm d,thin}$ and $M_{\rm d,thick}$ are the masses of the thin and thick disks, respectively \citep{McMillan2017}.

\subsection{Our Model for the Hot Gaseous Halo}
\label{initialdensity}

The spatial distribution of the hot gas in the MW's halo is quite complicated, affected by the gravity of the dark matter halo, the gravity of the Galatic disk and bulge (see Section \ref{diskbulge}), stellar and AGN feedback processes. Based on these considerations, here we adopt a general model to describe the radial profile of the hot halo gas:
\begin{equation}
 \label{ourmodel}
 \rho(r)=\frac{\rho_0}{(r+r_1)^{\alpha_1}(r+r_2)^{\alpha_2}}
\end{equation}
Here, $\rho_0$ is a constant normalized by the total gas mass within the virial radius $r_{vir}=206 $ kpc:
\begin{equation}
 \label{gasmass}
 M_{\rm g}=\int_{r_{\rm min}}^{r_{\rm vir}} \int_{0}^{\pi} \rho(r,\theta) 2 \pi r^2\sin \theta dr d\theta {,}
\end{equation}
where $r_{\rm min}$ is the inner boundary of our calculation domain (see Section \ref{Setup}). The values of $M_{\rm g}$ in our simulations are chosen to be a fraction of the missing baryon mass of the MW $M_{\rm mbar}=10^{11} M_\odot$ (see Sec. 1), as listed in the second column of Table 1. Observational estimates of $M_{\rm g}$ depend on the adopted profile for the spatial gas density distribution, and range from less than $50\%$ of $M_{\rm mbar}$ \citep{MillerBregman2015} to all of $M_{\rm mbar}$ \citep{Fang2013}.

The nonzero value of $r_1$ assumes that there is an inner thermal core in the gas distribution, as found in cosmological simulations, and here in all our simulations except for run 5, we adopt  $r_1 = \frac{3}{4}r_s$ \citep{MallerBullock2004}. $r_{2}$ represents the impact of stellar and AGN feedback processes on the halo gas distribution ($r_{2}>r_{\rm s}$), and here we assume that $r_{2}=100$ kpc or larger. $\alpha_1$ and $\alpha_2$ also affects the density profile, and to ensure that the gas density distribution approaches to the NFW distribution $\rho(r) \propto r^{-3}$ at very large radii, we assume that $\alpha_1 + \alpha_2 =3$.

\begin{table}
	\centering
	\begin{minipage}{75mm}
		\caption{List of Simulations}
		\label{simulatable}
		\begin{tabular}{@{}lcccccc}
			\hline
			\hline
			&$M_{g}$ & $r_1$ \footnote{$r_1$ = 23.4 kpc corresponds to $r_1$ = $\frac{3}{4}r_s$, which is adopted from \citet{MallerBullock2004}}
			&$r_2$
			& $\alpha_1$ \footnote{The value of $\alpha_2$ can be derived from $\alpha_1$ according to $\alpha_1 + \alpha_2 =3$.}
			& metallicity\\
			run & ($10^{11}M_\odot$)&(kpc)&(kpc)&&$Z_\odot$\\
			\hline
			1.............
			&1.0 & 23.4 & 100 & 1 & 0.3\\
			2.............
			&0.5 & 23.4 & 100 & 1 & 0.3\\
			3.............
			&0.3 & 23.4 & 100 & 1 & 0.3\\
			4.............
			&0.1 & 23.4 & 100 & 1 & 0.3\\
			5.............
			&1.0 & 70.0 & 100 & 1 & 0.3\\
			6.............
			&1.0 & 23.4 & 200 & 1 & 0.3\\
			7.............
			&1.0 & 23.4 & 500 & 1 & 0.3\\
			8.............
			&1.0 & 23.4 & 5000 & 1 & 0.3\\
			9.............
			&1.0 & 23.4 & 100 & 2 & 0.3\\
			10...........
			&1.0 & 23.4 & 100 & 1 & 0.1 \\
			11...........
			&1.0 & 23.4 & 100 & 1 & 1.0 \\
			\hline

		\end{tabular}
	\end{minipage}
\end{table}

Gas metallicity affects the cooling rate of the hot gas at about $10^{6}$ K strongly, and its impact on the thermodynamic evolution of the halo gas is investigated quantitatively in our simulations (see Section 3.4). Observational constraints on gas metallicity depend critically on the detection method and the adopted model, and are not firmly established yet. From observations of several high-velocity clouds (HVCs) and the Magellanic Stream, \citet{Gibson2000}, \citet{vanWoerdenWakker2004}, and \citet{Fox2005} imply that the halo gas metallicity may be predominantly sub-solar. \citet{MillerBregman2015} indicate that the halo gas has a sub-solar metallicity $Z\gtrsim$ 0.3$Z_\odot$ which decreases with radius. In the outer parts of the MW's halo, the gas metallicity may drop to $Z\sim 0.1Z_\odot$ \citep{Troitsky2017}. Here we adopted a spatially constant gas metallicity for simplicity, and investigated the impact of gas metallicity in our simulations with three different values of $Z$: $0.1Z_\odot$, $0.3Z_\odot$, and $Z_\odot$. To fully investigate how the properties of the halo gas affect the development of cooling flows, we performed a large suite of hydrodynamic simulations with different model parameters, as listed in Table \ref{simulatable}.

\subsection{Simulation Setup}
\label{Setup}

Assuming axisymmetry around the MW's rotational axis, we solve Equations (\ref{hydro1})-(\ref{hydro3}) in spherical coordinates $(r, \theta)$ with the ZEUS-MP code (\citealt{Hayes2006}). ZEUS-MP is a multi-physics, massively parallel, message-passing implementation (MPI) of the ZEUS code. Our computational domain extends from an inner boundary at $r_{\rm min}=1$ kpc to an outer boundary at $r_{\rm max}=250$ kpc in the radial direction. We have also performed several simulations with a much larger radial domain and found that our results are very robust. We adopt a non-uniform grid with $(\Delta r)_{i+1}/(\Delta r)_{i}=1.01745$ in the radial direction and a uniform grid in the angular direction within the computational domain $0\leq \theta \leq \pi$. The numbers of active meshes are $(N_r, N_\theta)=(324,180)$. We choose the outflow boundary condition at both the inner and outer radial boundaries where the gas is only allowed to flow out of our computational domain. For the boundaries in the angular direction at the poles, we choose the axis of symmetry boundary condition.

\begin{figure*}
  \centering
 \plottwo{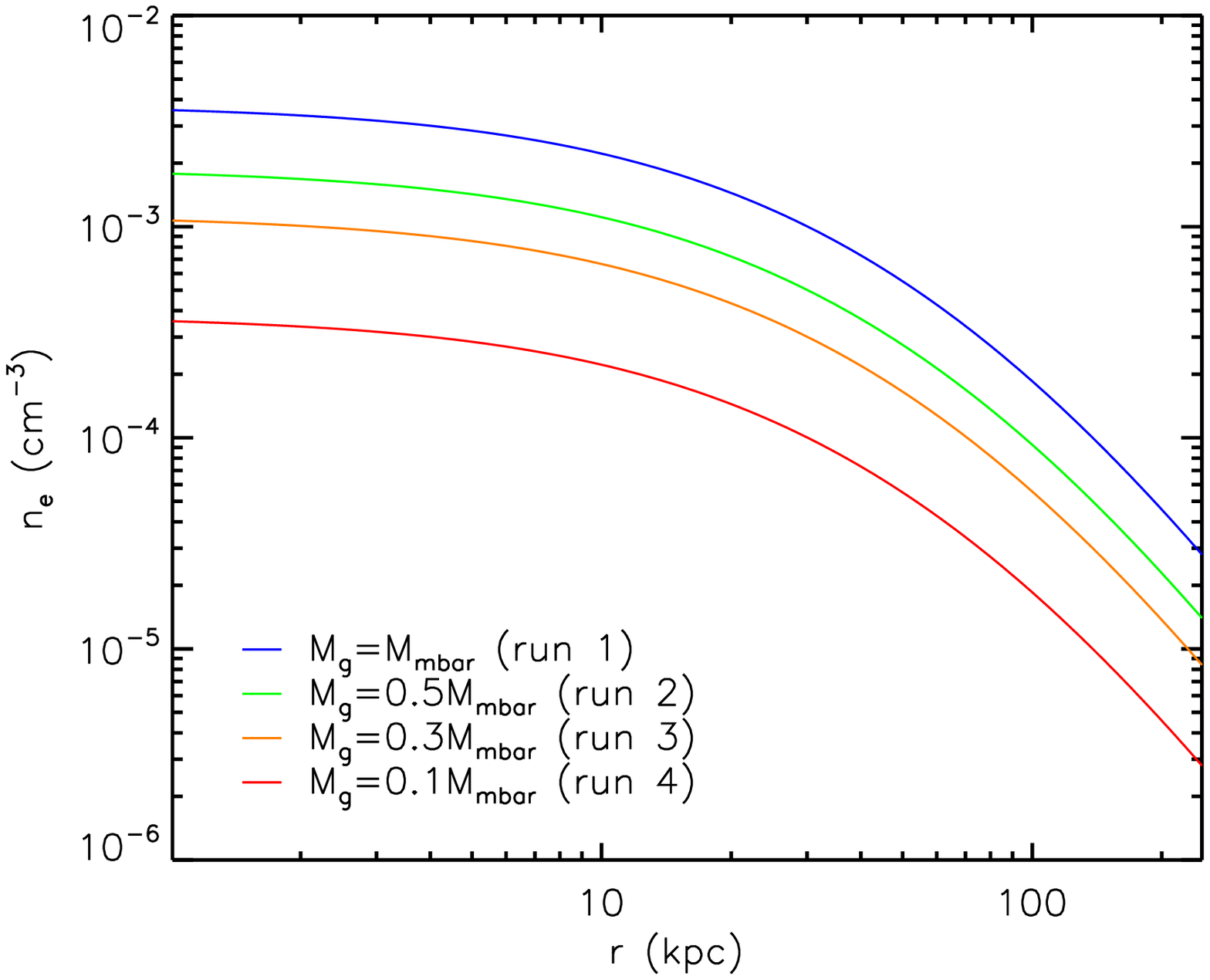}{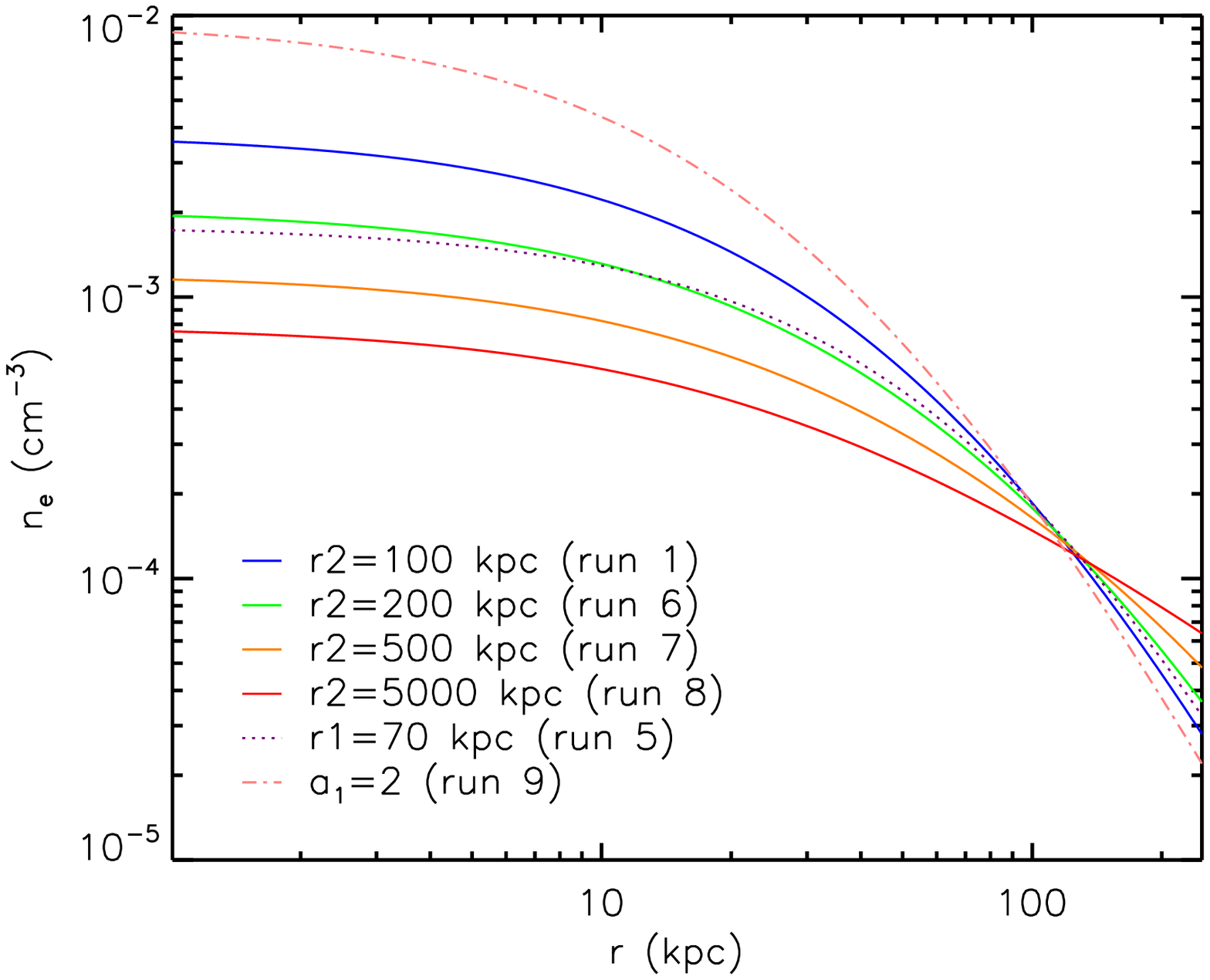}
 \caption{Initial gas density profiles in our simulations as listed in Table \ref{simulatable}. The legend in each panel shows the values of the parameters that are different from our fiducial model (run 1). $M_{\rm mbar}$ in the left panel denotes the total mass of the missing baryons in the MW, $M_{mbar}=1.0\times 10^{11}M_\odot$.}
 \label{plot1}
\end{figure*}

\begin{figure}
  \includegraphics[width=0.45\textwidth]{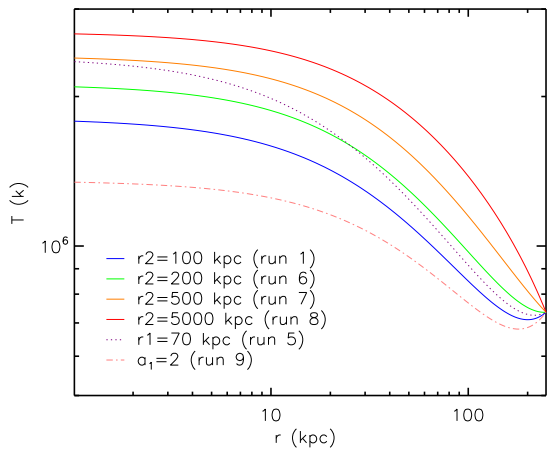}
  \caption{Initial gas temperature profiles in our simulations with different values of $r1$, $r_2$ and $\alpha_1$. Note that the temperature distribution is not affected by the total gas mass $M_{\rm g}$ and gas metallicity.}
  \label{plot2}
\end{figure}

\subsection{Initial Conditions}

We assume that the hot gas in the MW's halo is initially in hydrostatic equilibrium state, i.e., $\rho \nabla \Phi=-\nabla P $. 
The initial gas density distribution is described by Equation \ref{ourmodel} and discussed in Section \ref{initialdensity}. Figure \ref{plot1} shows the initial density profiles in our simulations with different model parameters as listed in Table \ref{simulatable}, including the total gas mass $M_{\rm g}$, $r_1$, $r_2$ and $\alpha_1$. We choose run 1 as our fiducial model, and the legend in each panel shows the values of the parameters that are different from run 1.

We assume that the initial gas temperature at the outer boundary $r=250$ kpc is $T_{\rm out}=7\times 10^5$ K, which is consistent with the result of the MB model shown in \citet{Fang2013}. The whole initial temperature profile can then be self-consistently derived from hydrostatic equilibrium and the initial density profile. With hydrostatic equilibrium, the gas temperature profile is determined by the density profile and the gravitational potential well. The normalization of the density profile ($M_{\rm g}$) and gas metallicity does not affect the derived temperature profile, which is shown in Figure \ref{plot2} for some of our simulations. We have also experimented with different values of $T_{\rm out}$ between $10^5 - 10^6$ K  that affects the initial temperature profile, and found that our main results presented in the following section are not affected substantially.

\section{RESULTS}
\label{section:results}

In this section, we present our results from a suite of hydrodynamic simulations with different model parameters. These simulations follow the thermodynamic evolution and the development of cooling flows in the hot gaseous halo of the MW. In Section \ref{Evolution}, we describe the evolution of the hot halo gas in our fiducial model (run 1). The impact of the total gas mass, the initial gas density distribution, and gas metallicity on our results are presented in Section \ref{Gas mass}, Section \ref{structure} and Section \ref{metallicity}, respectively. We discuss the impact of the Galactic disk and bulge in Section \ref{diskbulge}.

\subsection{Cooling Flows in Our Fiducial Run}
\label{Evolution}

\begin{figure*}
 \centering
\plottwo{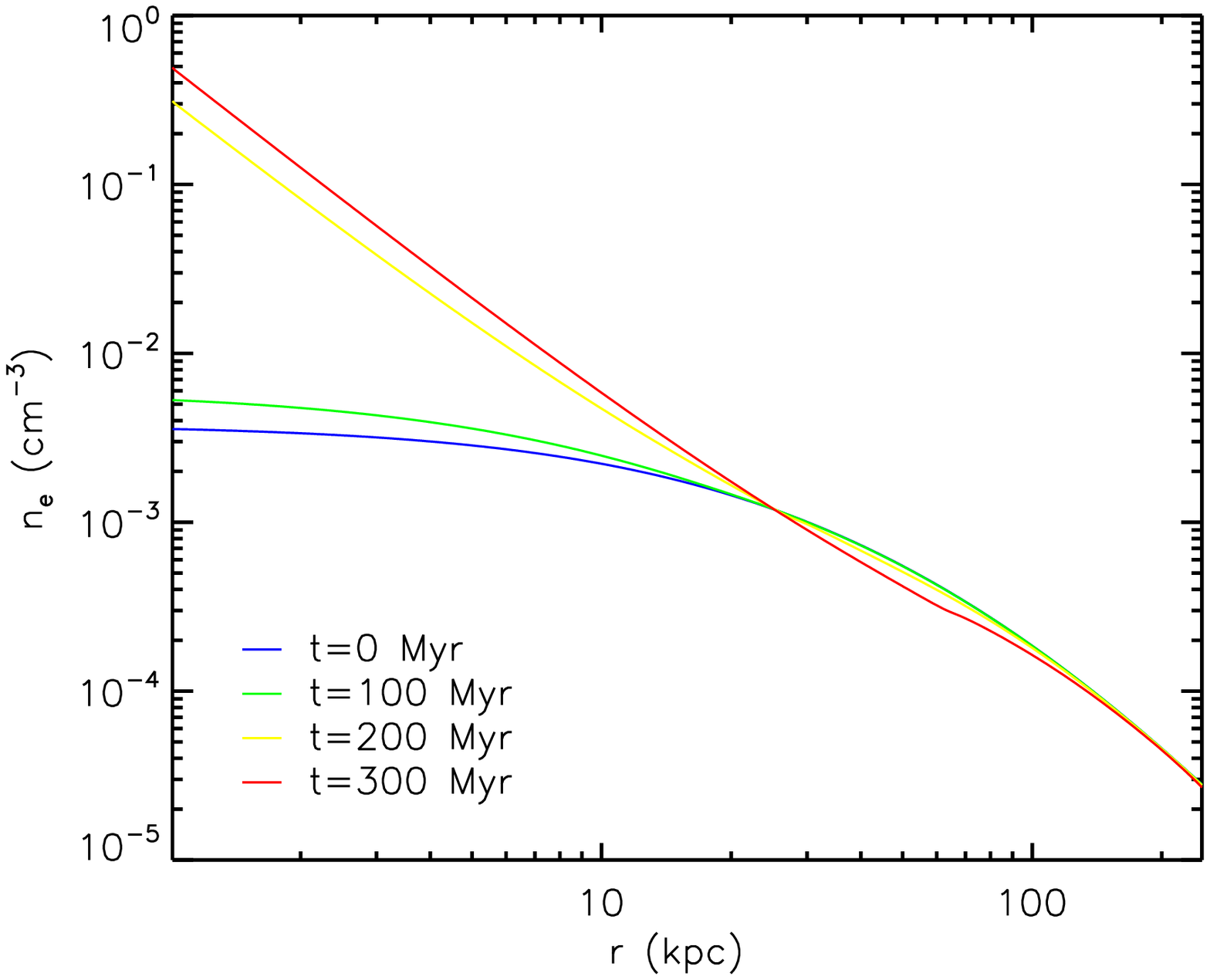}{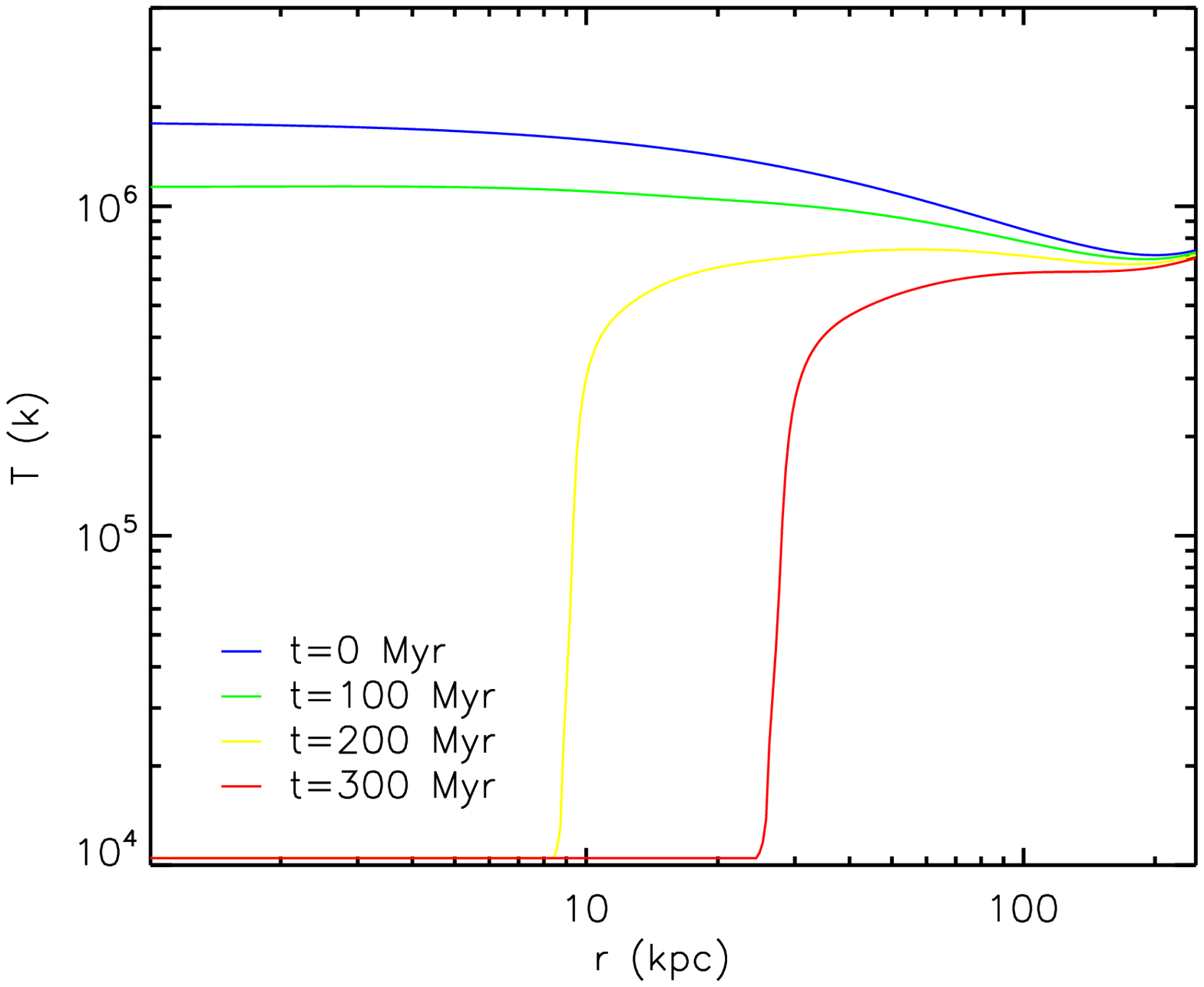}	
\caption{Evolution of the electron number density (left) and temperature (right) profiles of the hot gas in run 1. Note that a cooling catastrophe happens between $t=100$ and $200$ Myr.}
\label{plot3}
\end{figure*}

Here we use our fiducial model (run 1) to show the basic picture of the development of cooling flows in  the MW's halo. The simulation starts from hydrostatic equilibrium, and the evolution of the gas density and temperature profiles is shown in Figure \ref{plot3}. From $t=0$ Myr (blue line) to $t=100$ Myr (green line), the gas temperature in inner regions decreases slowly due to radiative cooling, and correspondingly, the gas density increases slowly. A cooling catastrophe happens sometime between $t=100$ and $200$ Myr, when the gas temperature within the inner several kpc drops dramatically from $\sim 10^6$ K to  $10^4$ K, which is the lower temperature floor that we set artificially in our simulations. The cooling catastrophe leads to gas inflows, resulting in a dramatic increase in the gas density in this inner region. At galactocentric distances of several tens kpc, the gas density drops slightly as the gas flows to inner regions after the cooling catastrophe. The time when the cooling catastrophe happens is roughly consistent with the initial cooling time of the gas in inner regions. 

When $t>$ 200 Myr, the gas in inner regions completely cools down, forming a cool core, and these cooled gas may become the raw material for future star formation activities in the Galaxy. At later times, the spatial size of the central cool core slowly grows with time. This whole picture is consistent with the development of spherically-symmetric cooling flows in galaxy clusters in the absence of any heating sources (e.g., see \citealt{guo18}). 

\subsection{Impact of the Halo Gas Mass}
\label{Gas mass}

\begin{figure}
 \includegraphics[width=0.5\textwidth]{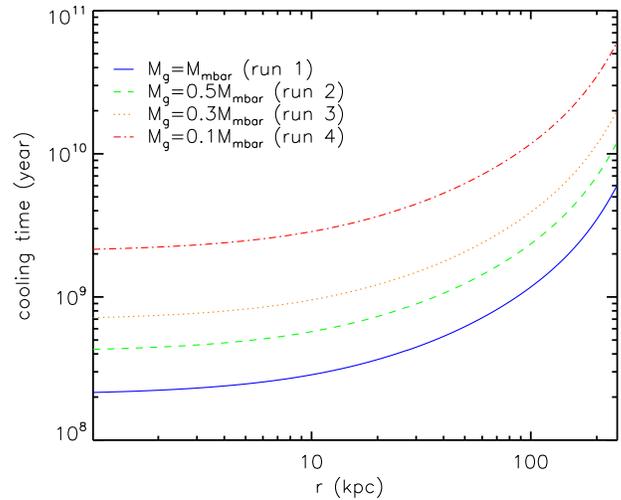}
 \caption{Cooling timescales of the hot gas in four of our simulations with different values of the initial halo gas mass $M_{\rm g}$. The blue solid, green dashed, orange dotted and red dot-dashed lines correspond to runs 1, 2, 3 and 4 with the corresponding total gas masses $M_{\rm g}=M_{\rm mbar}$, $0.5 M_{\rm mbar}$, $0.3 M_{\rm mbar}$ and $0.1 M_{\rm mbar}$, respectively.}
 \label{plot4}
\end{figure}

The cooling rate of the hot gas depends sensitively on the gas density, and the development of cooling flows is also expected to depend on it. The normalization of the gas density profile is described by the total halo gas mass $M_{\rm g}$ in our model, and here we investigate its impact on the gas evolution. Figure \ref{plot4} shows the initial cooling timescales $t_{\rm cool}=e/\mathcal{C}$ of the hot gas in four of our simulations with different total gas masses. As expected, the cooling time increases with radius, as the gas density decreases outward. At each radius, the gas cooling time decreases as the value of $M_{\rm g}$ increases. When $M_{\rm g}=0.1\times M_{\rm mbar}$ in run 4, the gas cooling time exceeds 2 Gyr even in inner regions, indicating that the central cooling catastrophe would not develop within about $2$ Gyr. When $M_{\rm g}= M_{\rm mbar}$, the cooling time of the hot gas in inner regions is about 200 Myr, consistent with the time when the cooling catastrophe happens in run 1 (see Section \ref{Evolution}).

\begin{figure}
 \includegraphics[width=0.5\textwidth]{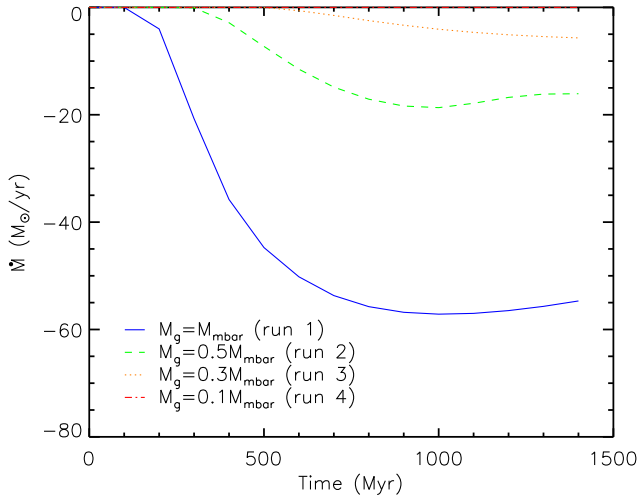}
 \caption{Time evolution of the mass inflow rates across the inner boundary in runs 1, 2, 3, and 4 with different halo gas masses. Negative values of $\dot{M}$ indicate inflows.}
 \label{plot5}
\end{figure}

To quantify the strength of the developed cooling flows, we show the evolution of the mass inflow rates in runs 1, 2, 3, and 4 in Figure \ref{plot5}. Here the mass inflow rate $\dot{M}(t)$ is evaluated at the inner boundary $r_{\rm min}=1$ kpc in our simulations,
\begin{equation}
 \dot{M}(t) = 2\pi r_{\rm min}^2 \int_{0}^{\pi} \rho(\theta,t)v_r(\theta,t)\sin \theta d\theta {.}
\end{equation}
As shown in Figure \ref{plot5}, the mass inflow rate evolves slowly before the cooling catastrophe. After the cooling catastrophe develops, the mass inflow rate increases gradually and eventually reaches to a quasi-steady state. At very late times, the mass inflow rate decreases slightly as the total gas mass within our computational domain drops due to gas inflows across the inner boundary (note that gas inflows across the outer boundary are prohibited by our outflow boundary condition).

It is clear from Figure \ref{plot5} that the total mass of the halo gas has a great impact on the mass inflow rate. The higher the halo gas mass, the larger the mass inflow rate. When $M_{\rm g}=M_{\rm mbar}$, the mass inflow rate in the quasi-steady state approaches $50-60~M_{\odot}$ yr$^{-1}$. Even when $M_{\rm g}$ drops to $0.3M_{\rm mbar}$ as in run 3, the final mass inflow rate is still about $5~M_{\odot}$ yr$^{-1}$. These inflows bring cold gas to central regions of the Galaxy, and could thus significantly enhance the star formation rate (SFR) in the MW. If the gas mass in the Galactic halo is very low, say, only about one tenth of the missing baryons in the MW as in run 4, the gas cooling time is then very longer ($>2$ Gyr; see Figure \ref{plot4}), and the gas inflow rate at around 1 Gyr only reaches about $\dot{M}\sim 10^{-5} M_\odot$ yr$^{-1}$.

It is widely believed that the current observed SFR in the MW is about $1-2$ $M_{\odot}$ yr$^{-1}$ (\citealt{RobitailleWhitney2010}; \citealt{chomiuk11}). Recent observations suggest that the bulk of the stars at the GC were formed at least 8 Gyr ago, and the star formation activity there was very quiescent during most times of the past 8 Gyr (\citealt{Nogueras2019}). The low SFR in the MW and particularly the long-term low level of star formation activity at the GC indicate that there must be additional heating sources in the MW to suppress cooling flows, unless the total mass of the halo gas is very small $M_{\rm g}\lesssim 0.1M_{\rm mbar}$, which results in inefficient gas cooling.

\subsection{Impact of the Gas Density Distribution}
\label{structure}

$M_{\rm g}$ determines the total mass of the halo gas, while the gas density distribution is also affected by the parameters $r_1$, $r_2$, $\alpha_1$, and $\alpha_2$, as shown in Equation (\ref{ourmodel}). Here $r_1<r_2$ and $\alpha_2=3-\alpha_1$. The values of these parameters determine where the halo gas with the total mass $M_{\rm g}$ is distributed radially. The density profile is flat at $r \ll r_1$, scales roughly as $r^{-\alpha_1}$ at $r_1 \ll r \ll r_2$, and as $r^{-3}$ at $ r \gg r_2$. In this subsection we discuss the impact of the halo structure parameters $r_1$, $r_2$, and $\alpha_1$ on the evolution of the halo gas.

\begin{figure*}
	\centering
	\plottwo{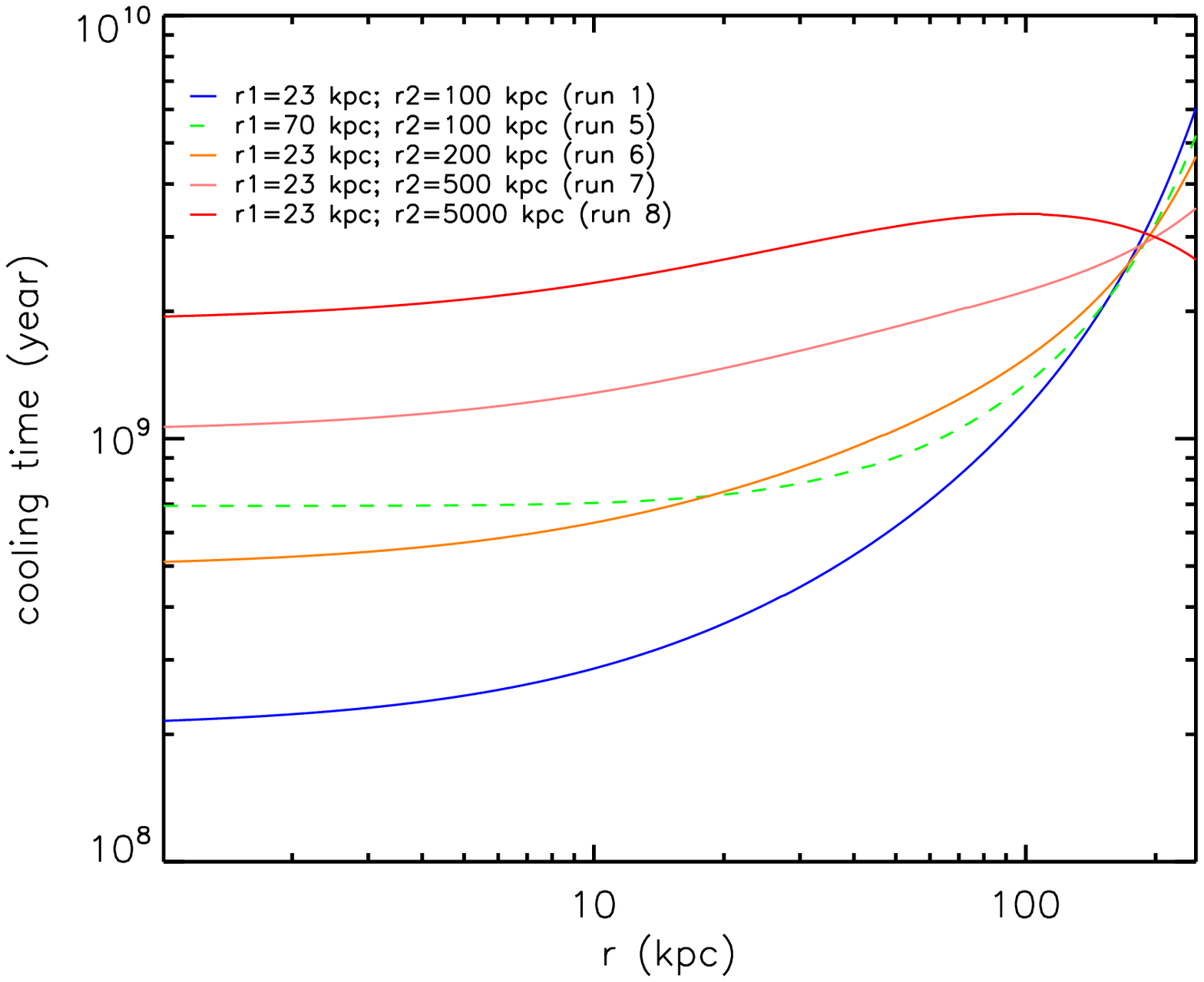}{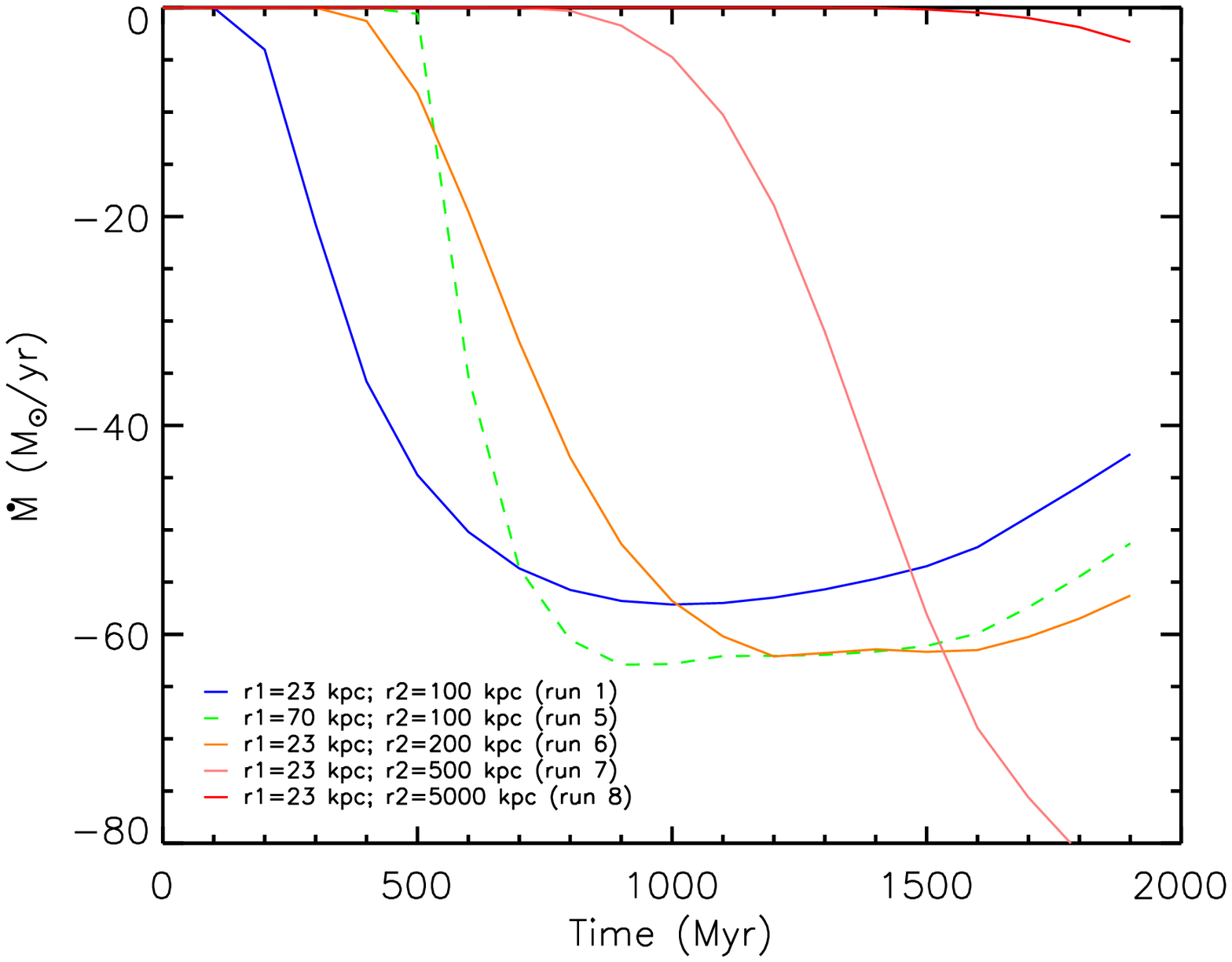}
	\caption{Initial gas cooling timescales as a function of radius (left panel) and the temporal evolution of the mass inflow rate across the inner boundary (right panel) in some of our simulations with different values of the model parameters $r_1$ and $r_2$.}
	\label{plot6}
\end{figure*}

Figure \ref{plot6} shows the initial gas cooling timescales and the evolution of the mass inflow rates in a series of five simulations with different values of $r_1$ and $r_2$. Note that the halo gas mass within $r_{\rm vir}$ is the same in these runs, and $r_1$ and $r_2$ affect where the halo gas is distributed spatially. When the value of $r_1$ increases from $23$ kpc in run 1 to $70$ kpc in run 5, the inner thermal core expands, resulting in a decrease in the central gas density (see Figure \ref{plot1}) and an increase in the central gas cooling time (see the left panel of Figure \ref{plot6}). Similarly, as the value of $r_2$ increases, the halo gas is distributed more extendedly, resulting in a decrease in the central gas density (see Figure \ref{plot1}) and an increase in the central gas cooling time (see the left panel of Figure \ref{plot6}). Therefore, the cooling catastrophe happens at a later time when the value of  $r_1$ or $r_2$ is larger, as clearly shown in the right panel of Figure \ref{plot6}.

\begin{figure*}
	\centering
	\plottwo{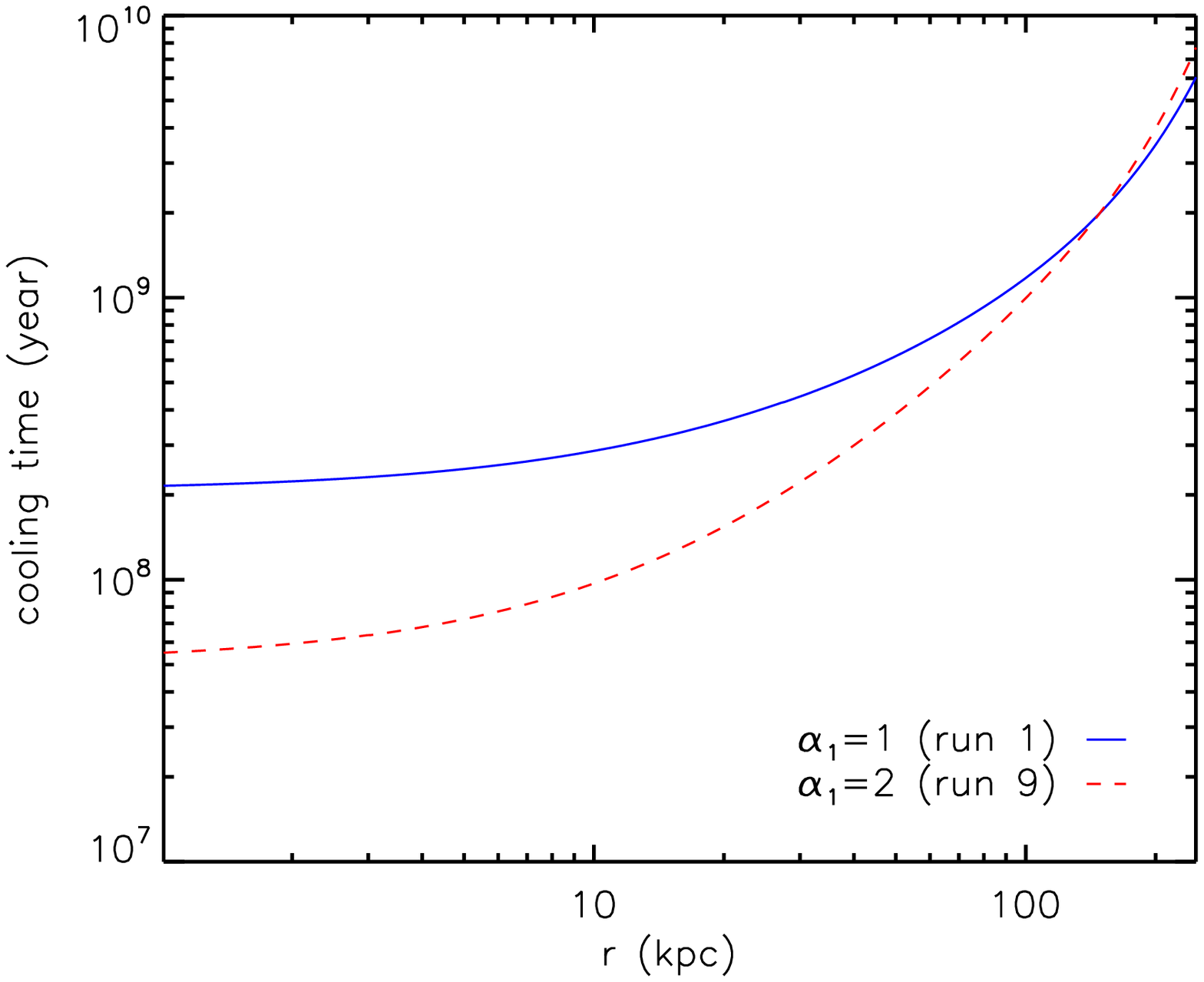}{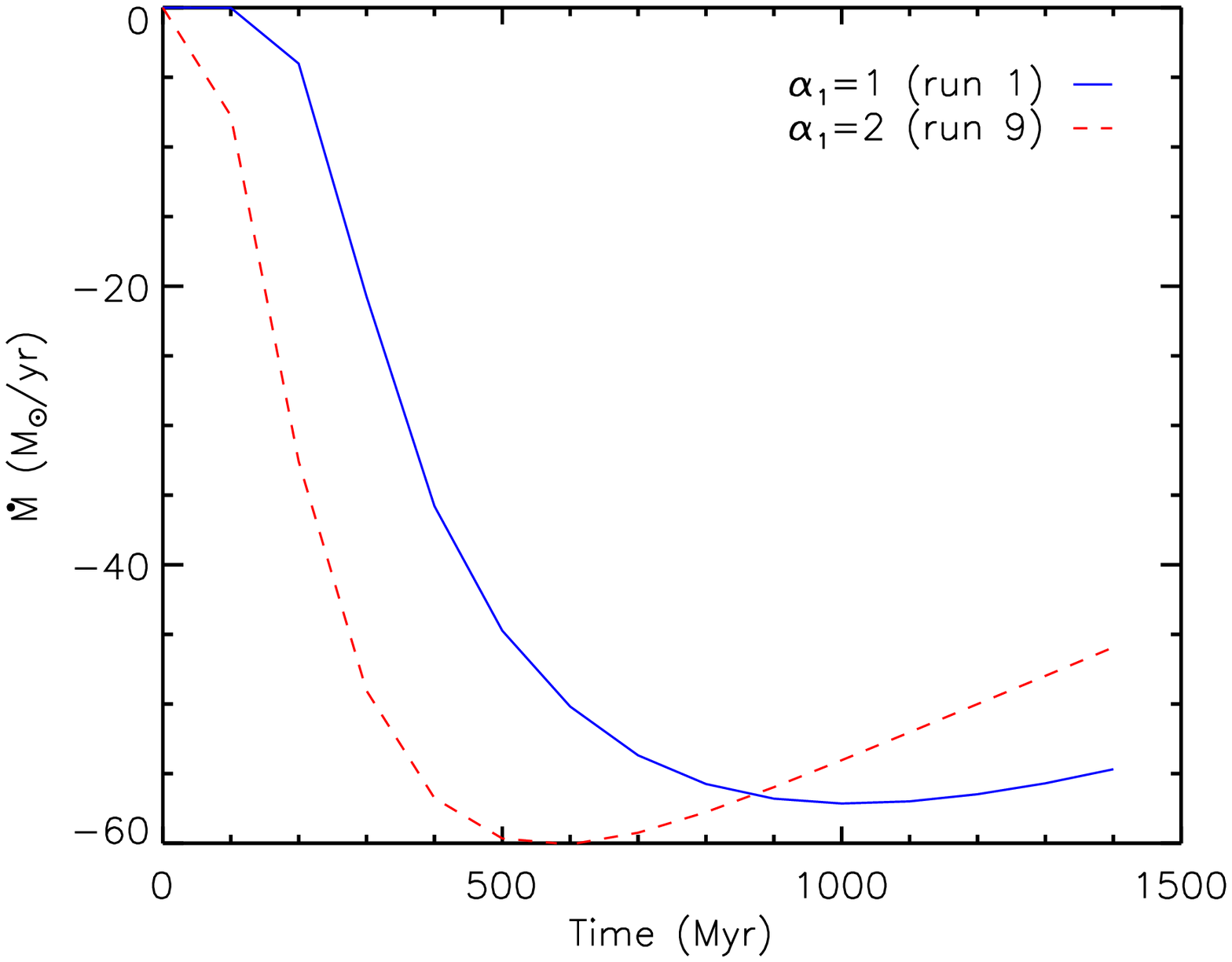}
	\caption{Initial gas cooling timescales as a function of radius (left panel) and the temporal evolution of the mass inflow rate across the inner boundary (right panel) in some of our simulations with different values of the model parameters $\alpha_{1}$ and $\alpha_2$. Note that $\alpha_2=3-\alpha_1$.}
	\label{plot7}
\end{figure*}

Figure \ref{plot7} shows the initial gas cooling timescales and the evolution of the mass inflow rate in two simulations (runs 1 and 9) with different values of $\alpha_1$. As $\alpha_1$ increases, the inner density core becomes stronger, resulting in higher central gas densities and shorter central gas cooling times. Thus the cooling catastrophe develops at an earlier time in run 9 with $\alpha_1=2$ than in run 1 with $\alpha_1=1$.

In summary, for a fixed halo gas mass, the gas density distribution affects the time when the cooling catastrophe happens. More extendedly the halo gas is distributed, later the cooling catastrophe starts. However, it is remarkable that the mass inflow rate after the cooling flow reaches the quasi-steady state does not depend sensitively on the values of $r_1$, $r_2$, and $\alpha_1$, as shown in the right panels of Figures \ref{plot6} and \ref{plot7}. On the other hand, as discussed in Section 3.2, the mass inflow rate depends strongly on the total halo gas mass. We note that the gas distribution in run 8 is very extended with $r_{2}=5000$ kpc, approaching the MB distribution, which results in a very long central gas cooling time, and therefore the cooling catastrophe has not yet started at $t<2$ Gyr, which explains the low mass inflow rates at $t < 2$ Gyr shown in Figure \ref{plot6}.

\subsection{Impact of Gas Metallicity}
\label{metallicity}

\begin{figure*}
	\centering
	\plottwo{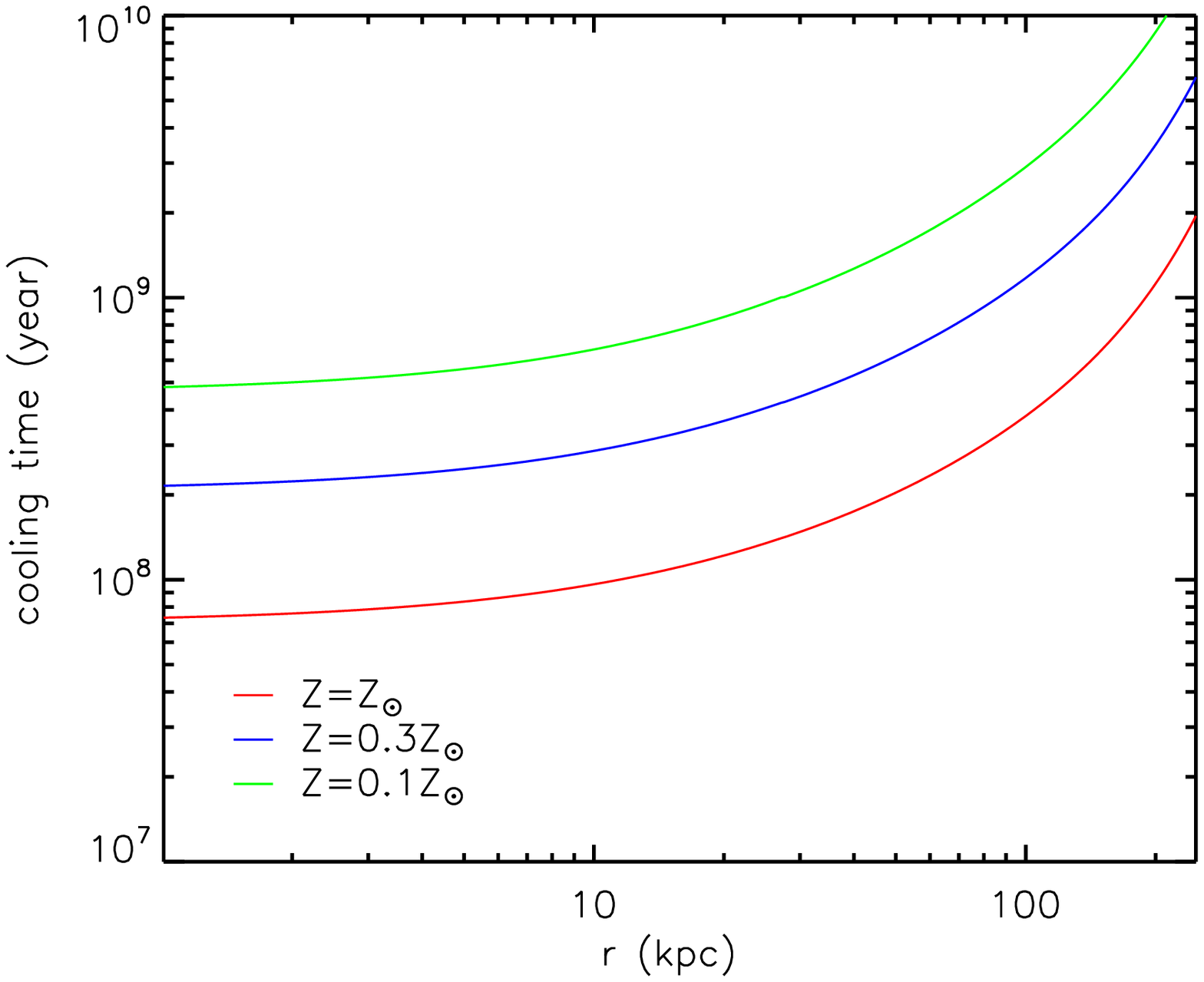}{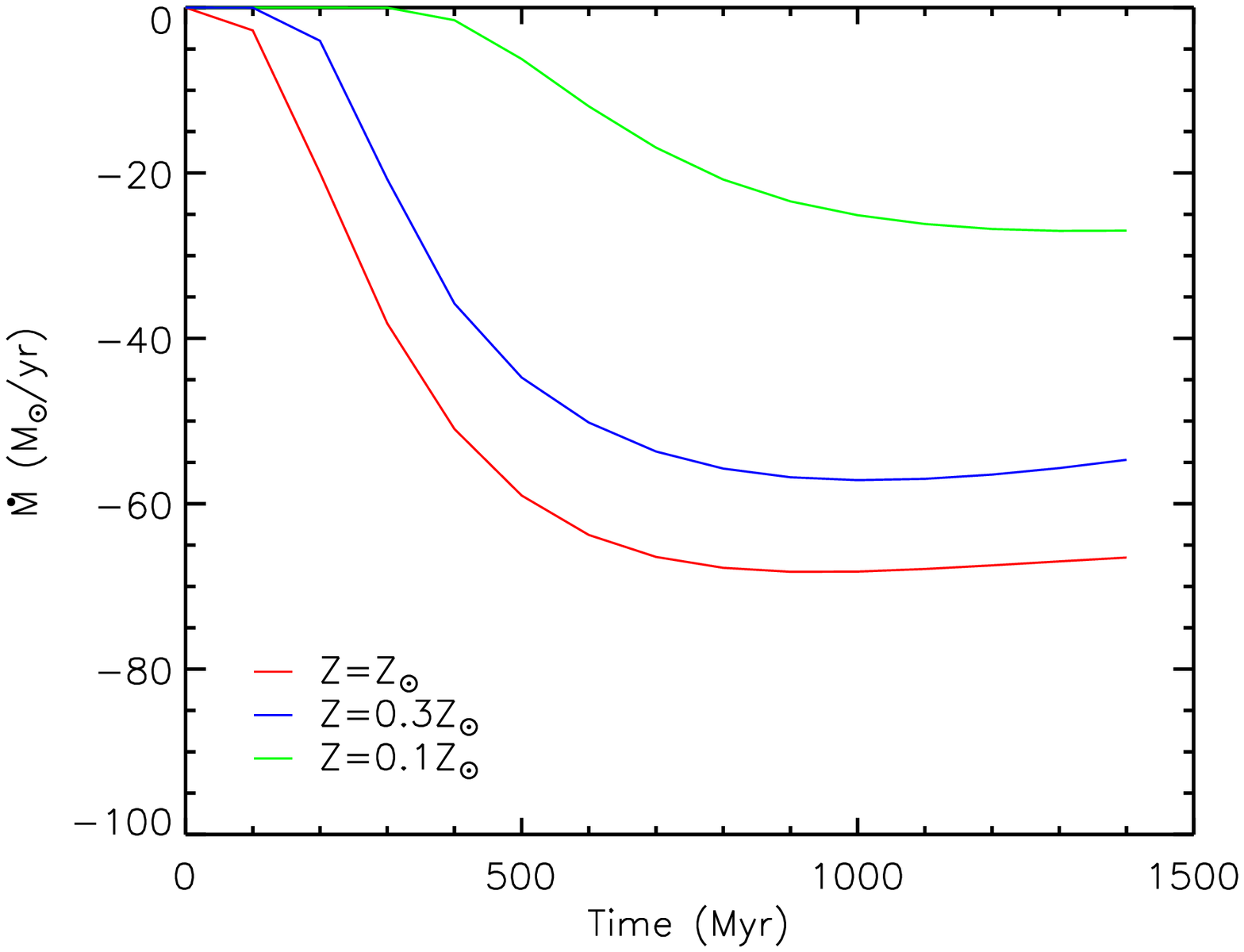}
	\caption{Initial gas cooling timescales as a function of radius (left panel) and the temporal evolution of the mass inflow rate across the inner boundary (right panel) in our fiducial model (run 1) with different gas metallicities.}
	\label{plot8}
\end{figure*}

Here we investigate the impact on the evolution of the halo gas from gas metallicity, which significantly affects the gas cooling rate. We present our results from our fiducial run with three different values of gas metallicity $Z=0.1Z_\odot$, $0.3Z_\odot$, and $Z_\odot$ in Figure \ref{plot8}, which shows the initial gas cooling timescales (left panel) and the temporal evolution of the mass inflow rate across the inner boundary in these three runs. It is clear that the gas cooling time decreases with increasing metallicity. With higher gas metallicities, the cooling catastrophe starts earlier, and the mass inflow rate in the quasi-steady state is also larger. With $M_{\rm g}= M_{\rm mbar}$ in run 1, the final mass inflow rate increases from about $25~ M_\odot$ yr$^{-1}$ when $Z=0.1Z_\odot$ to about $70 ~M_\odot$ yr$^{-1}$ when $Z=Z_\odot$. Thus, if $M_{\rm g}= M_{\rm mbar}$ and $Z\gtrsim 0.1Z_\odot$, the predicted mass inflow rate is much larger than the observed SFR in the MW of about $1-2 M_\odot$ yr$^{-1}$ (\citealt{RobitailleWhitney2010}; \citealt{chomiuk11}), indicating that heatings from star formation or AGN activities may be important in the Galactic halo.

\subsection{Impact of the Galactic Disk and Bulge}
\label{diskbulge}

In all the simulations presented in the previous subsections, we only consider the gravity of the dark matter halo, while the contribution of the Galactic disk and bulge to the gravity is neglected. Here we investigate the impact of the gravity of the Galactic disk and bulge on the evolution of the cooling flows. Note that the gravitational potential of the Galactic disk is not spherically-symmetric (see Equation 12). We choose run 1 as our fiducial model and present two simulations with and without the gravity of the Galactic disk and bulge. We adopt the same spherically-symmetric initial gas density profile as in run 1, and solve the gas temperature distribution assuming hydrostatic equilibrium. 

\begin{figure}
  \includegraphics[width=0.5\textwidth]{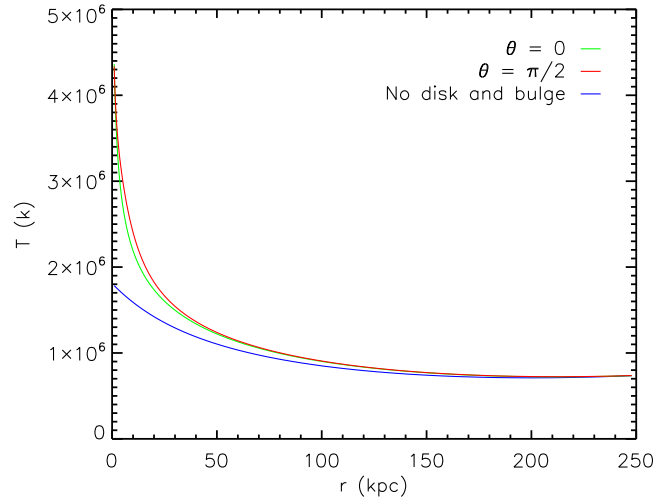}
  \caption{Initial temperature profiles as a function of radius along different angles: $\theta=0$ (green) and $\pi/2$ (red) in run 1 when the gravity of the Galactic disk and bulge is included. The temperature is solved from hydrostatic equilibrium with a spherically-symmetric gas density profile the same as in run 1. As a comparison, the blue line is the corresponding temperature profile in run 1 when the gravity of the Galactic disk and bulge is not included.}
  \label{plot9}
\end{figure}

\begin{figure*}
	\centering
	\plottwo{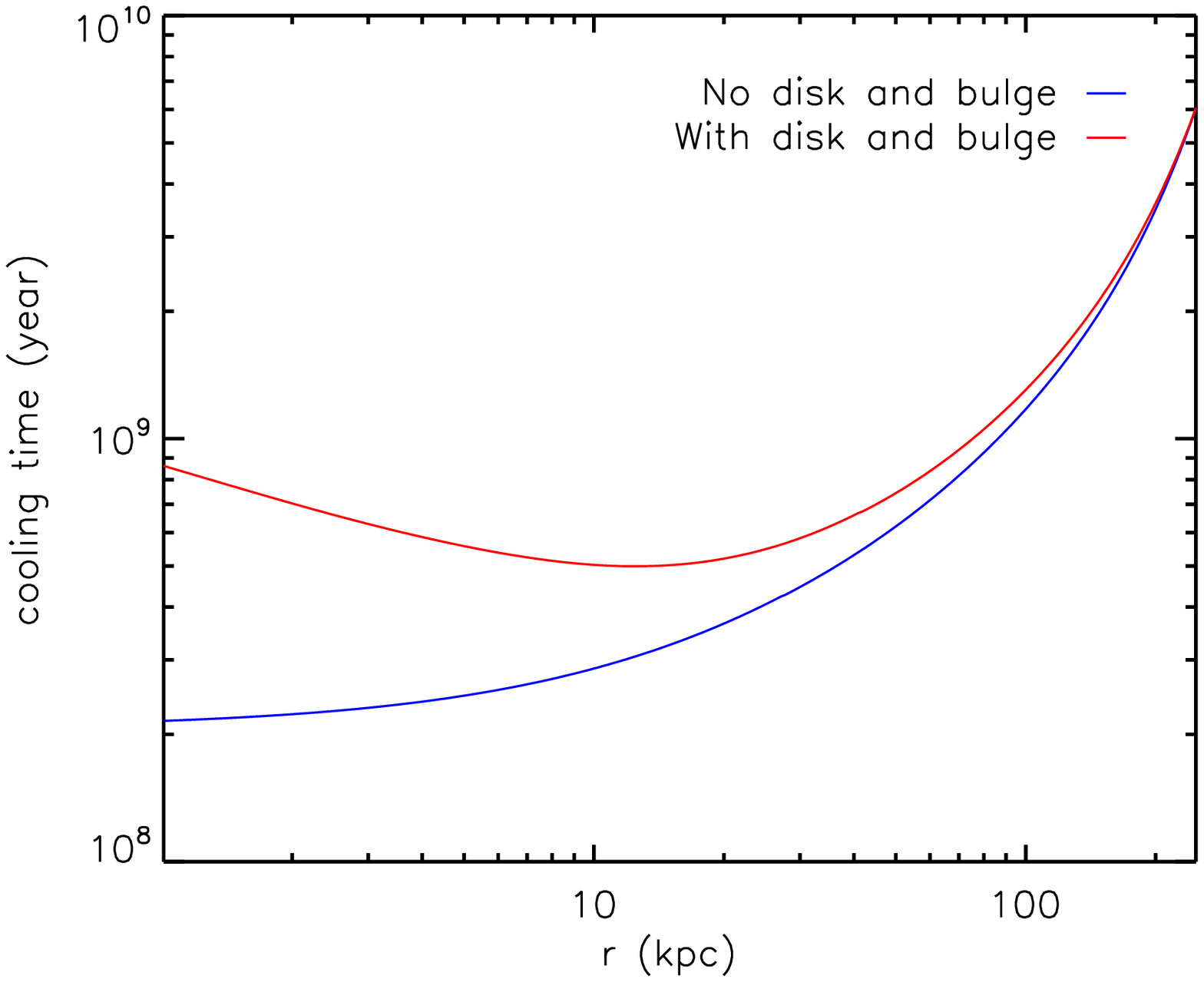}{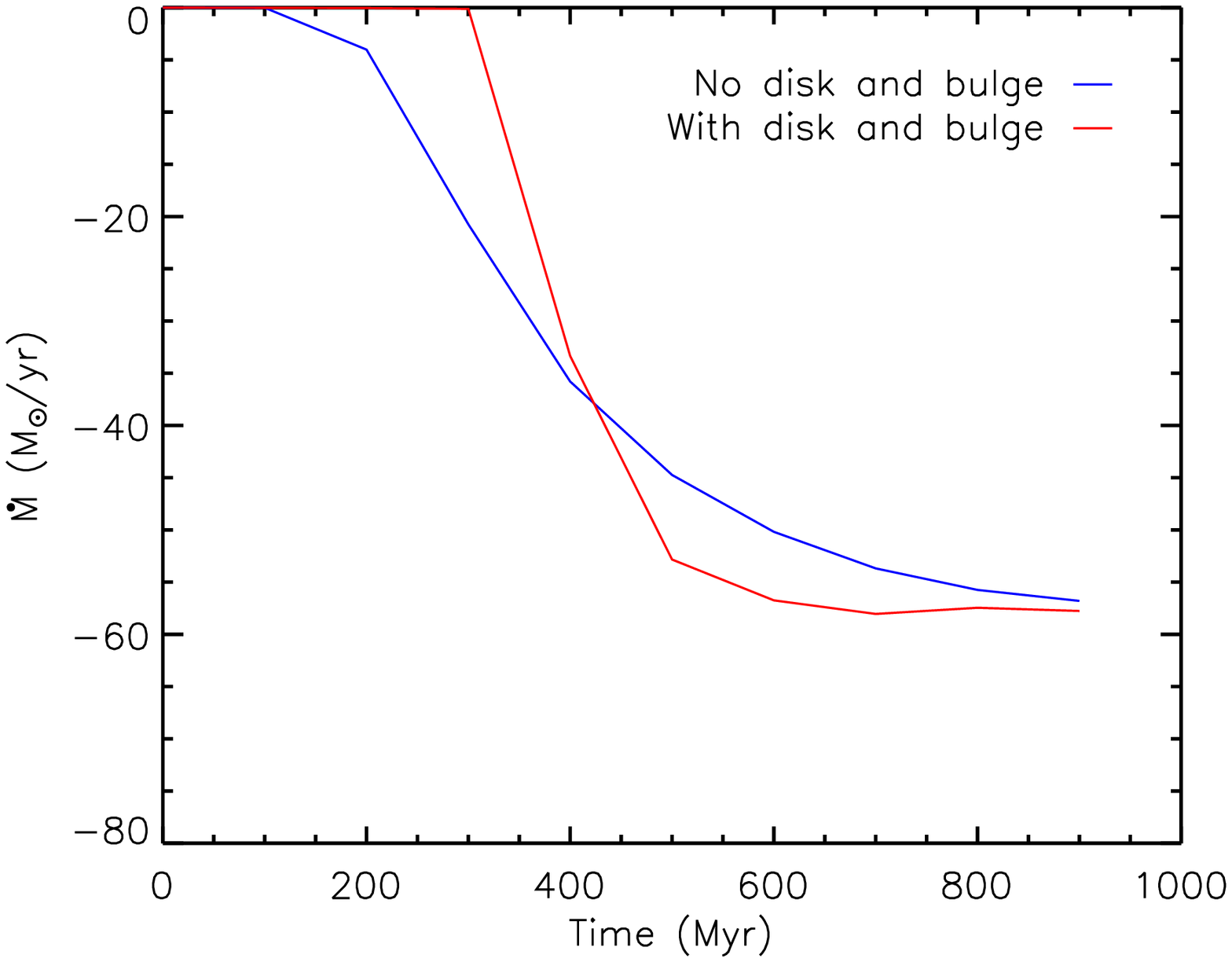}
	\caption{Impact of the gravity of the Galactic disk and bulge on the evolution of the cooling flows. We choose run 1 as our fiducial model and present two simulations with (red) and without (blue) the impact of the Galactic disk and bulge. The left panel shows the angle-averaged cooling timescale as a function of radius at $t=0$, while the right panel shows the temporal evolution of the mass inflow rate across the inner boundary.}
	\label{plot10}
\end{figure*}

Figure \ref{plot9} shows the initial temperature profiles along the Galactic rotation axis ($\theta=0$) and the Galactic plane ($\theta=\pi/2$). It is clear that the gas temperatures along these two angles are very close to each other at the same radius, and the gas temperature along the Galactic plane is slightly higher in inner regions due to stronger gravity. Comparing the simulations with and without the gravity of the disk and bulge, it is clear that the gravity of the disk and bulge mainly affects the inner region with $r\lesssim 50$ kpc. At $r \gtrsim 50$ kpc, the impact of the Galactic disk and bulge is very little, and the gas temperature profiles in these two cases are very close to each other. With the gravity of the disk and bulge, the gas temperatures in inner regions are higher, leading to longer cooling times there (see the left panel in Figure \ref{plot10}) and a later start time of the cooling catastrophe (see the right panel in Figure \ref{plot10}). However, the right panel in Figure \ref{plot10} clearly shows that the gravity of the Galactic disk and bulge does not substantially affect the mass inflow rate at the final quasi-steady state, which is mainly determined by the total gas mass in the halo and the gas metallicity.

In reality, the gravity from the Galactic disk and bulge affects both the density and temperature distributions of the halo gas in inner regions. We performed an additional simulation, showing that, if the initial gas temperature profile is fixed, the equilibrium gas density profile will be enhanced in the inner region ($r\lesssim 50$ kpc). For a fixed total mass of the halo gas, this means that the halo gas is distributed more centrally-peaked, leading to shorter central cooling times and an earlier start time of the cooling catastrophe. However, the mass inflow rate at the final quasi-steady state does not vary much compared to our fiducial run with the dark matter potential only. We thus conclude that the gravity from the Galactic bulge and disk affects gas properties in inner regions, but has little effect on the final mass inflow rate.

\section{comparison with other models}
\label{section:othermodels}

\begin{figure}
  \includegraphics[width=0.5\textwidth]{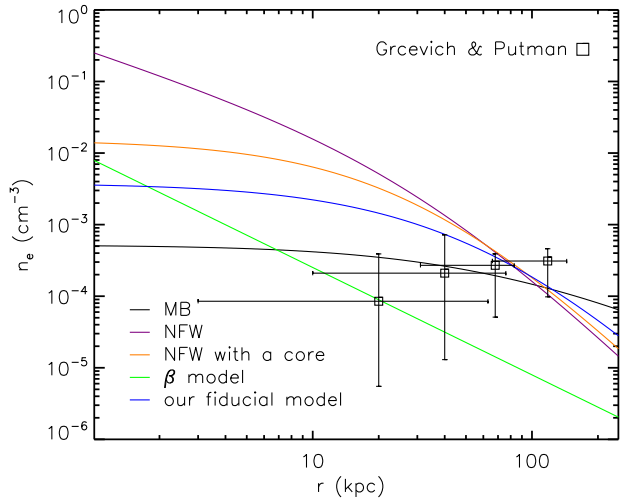}
  \caption{Radial profiles of electron number density in a variety of gas density models: the MB model, the NFW model, the cored-NFW model, the $\beta$ model, and our fiducial model (run 1). The squares are observational constraints adopted from \citet{GrcevichPutman2009}.}
  \label{plot11}
\end{figure}

The spatial distribution of the hot gas in the Galactic halo has not yet been well constrained by observations, and in our simulations, we use a new model (eqaution \ref{ourmodel}) for it. There are several widely used models for the halo gas distribution in the literature. Here in this section, we investigate four such models: the MB model, the NFW model, the cored-NFW model, and the $\beta$ model. In particular, we choose these density models as our initial conditions in the simulations, and investigate the development of cooling flows. We compare the main results of our density model with those of these four density models adopted from the literature. Here we first briefly describe these four models and the model parameters used in the simulations:

(1) The MB model assumes that the hot gas in the halo is isentropic with a polytropic index of $5/3$ and is in hydrostatic equilibrium within the gravitational potential of the MW's NFW dark matter halo (\citealt{MallerBullock2004}; \citealt{Fang2013}). The gas density profile can be described by \citep{MallerBullock2004}:
\begin{equation}
 \rho_{\rm MB}(x)=\rho_{\rm m}[1+\frac{3.7}{x}\ln(1+x)-\frac{3.7}{C_v}]^{3/2} {~,}
\end{equation}
where $x\equiv r/r_s$. Following \citet{Fang2013}, we determine the normalization parameter $\rho_{\rm m}$ by requiring that the total hot gas mass within the virial radius equals to $M_{\rm mbar}=10^{11}~M_\odot$. The temperature profile is naturally derived from hydrostatic equilibrium as described in Section~\ref{initialdensity}, and is consistent with that presented in \citet{MallerBullock2004} and \citet{Fang2013}. 

(2) The NFW model assumes that the density distribution of the hot halo gas scales with the NFW dark matter density distribution with a reduction factor:
\begin{equation}
 \rho_{\rm NFW}(x)=\frac{\rho_{\rm n}}{x(1+x)^2}{~,}
\end{equation}
where $x$ is the same as before, $x \equiv r/r_s$, and the normalization parameter $\rho_{\rm n}$ is determined by the total hot gas mass within the virial radius $M_{\rm g}= M_{\rm mbar}$. As in our fiducial model, the temperature profile is determined by the assumption of hydrostatic equilibrium. Note that the NFW model for the halo gas in \citet{Fang2013} assumes a spatially constant temperature $T_{\rm h}$, and therefore, the hot halo gas is not strictly in a hydrostatic equilibrium state there. 

(3) The cored-NFW model assumes that the hot gas in the halo follows the NFW model except for an inner core, as suggested by a large suite of non-radiative cosmological gasdynamical simulations of galaxy clusters in \citet{Frenk1999}. Here we adopt the cored-NFW model for the density distribution of the hot halo gas according to \citet{MallerBullock2004}:
\begin{equation}
\rho_{\rm F}(r)=\frac{r_s^3\rho_{\rm f}}{[r+(3/4)r_s](r+r_s)^2}{~,}
\end{equation}
where $\rho_{\rm f}$ is the normalization parameter to describe the total gas mass within the virial radius, and we assume $M_{\rm g}= M_{\rm mbar}$ as in the above two models. The size of the inner thermal core is $3r_{s}/4$, as also adopted in most of our simulations (see Table 1). The temperature profile is also solved from the assumption of hydrostatic equilibrium. This model is very similar to the NFW model except for a flat inner core, and is essentially the same as our model if we choose $r_1=3r_{s}/4$ and $r_2=r_s$ in Equation \ref{ourmodel}. 

(4) The $\beta$ model is often adopted to interpret the observed X-ray surface brightness profiles of early-type (\citealt{Forman1985}, \citealt{O'Sullivan2003}) and late-type (\citealt{AndersonBregman2011}, \citealt{Dai2012}) galaxies. The classical $\beta$ model is described as:
\begin{equation}
\label{beta1}
 \rho_{\beta} (r)=\rho_{\rm b}(1+(r/r_c)^2)^{-3\beta/2}{~,}
\end{equation}
where $\rho_{\rm b}$ is the central density, $r_c$ is the core radius, and typically $r_c \lesssim 5$ kpc. Following \citet{MillerBregman2015}, we simplify the $\beta$ model to a simple power law as $r>r_c$ in most regions: 
\begin{equation}
 \label{beta2}
 \rho_{\beta}(r)\approx \frac{\rho_b r_c^{3\beta}}{r^{3\beta}}{~,}
\end{equation} 
We adopt the best-fit values of the model parameters $\beta=0.5$ and $\rho_b r_c^{3\beta}$ from \citet{MillerBregman2015}. With these parameters, the total gas mass within the virial radius only accounts for $6.5\%$ of the total missing baryon mass in the MW $M_{\rm mbar}$. For the gas temperature in the halo, we also follow \citet{MillerBregman2015} to set a spatially constant temperature $T_{\rm halo}=10^{6.3}$ K, and thus the hot gas in the halo is not strictly in hydrostatic equilibrium.

\begin{figure}
 \includegraphics[width=0.5\textwidth]{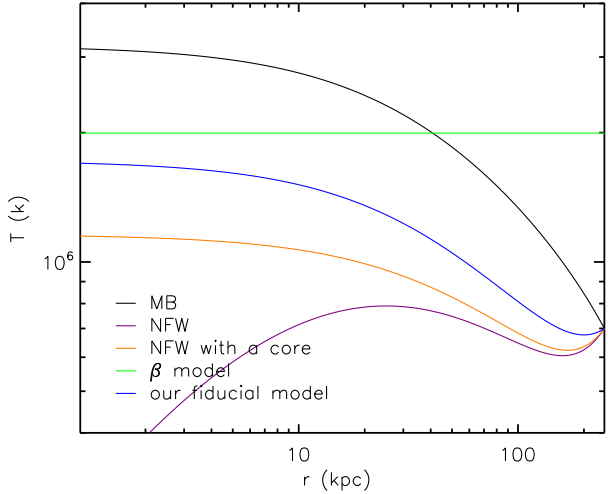}
 \caption{Corresponding temperature profiles as a function of radius for a variety of gas density models shown in Figure 11. See text in Section 4 for more details.}
 \label{plot12}
\end{figure}

\begin{figure}
 \includegraphics[width=0.5\textwidth]{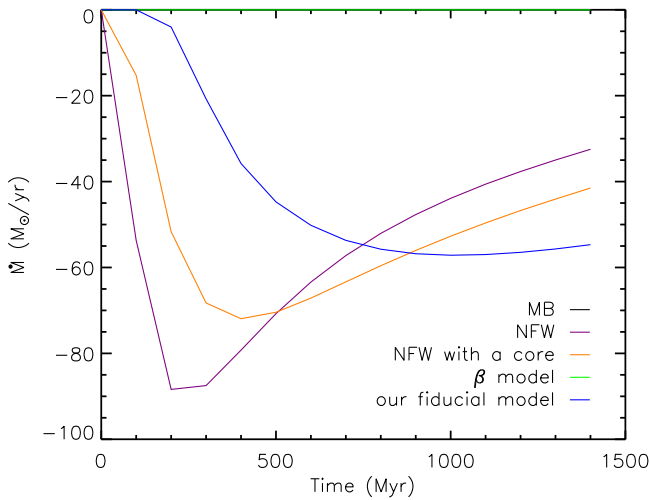}
 \caption{Temporal evolution of the mass inflow rates across the inner boundary in a variety of halo gas models shown in Figures 11 and 12.}
 \label{plot13}
\end{figure}

Figure \ref{plot11} shows the radial profiles of electron number density in the above four models and our fiducial model (the initial density profile in run 1). As a comparison, we also show density estimates from ram-pressure stripping arguments in \citet{GrcevichPutman2009}. It is clear that the gas density profile in our model lies between the centrally-peaked NFW model and the MB model, which is the most spatially extended model in these five density models. The corresponding temperature profiles of these five models are shown in Figure \ref{plot12}. It seems that the NFW model is unphysical for the hot gas in the Galactic halo as the temperature drops inward in inner regions and is lower than $10^6$ in most regions, while X-ray observations suggest that the halo gas temperature is around $\log T =6.1- 6.4$ (\citealt{Wang2005}, \citealt{Yao2007}, \citealt{Hagihara2010}).

The development of cooling flows in these four models is consistent with our results presented in Section 3. Figure \ref{plot13} shows the temporal evolution of the mass inflow rates across the inner boundary in these four models and in our fiducial model (run 1) for comparison. As in run 8 (see Figure 6), the gas in the MB model has very low densities in inner regions, resulting in very long cooling times ($\gtrsim 2$ Gyr). Thus the cooling catastrophe has not yet started during the simulation time of $t\sim 1.5$ Gyr in the simulation for the MB model, and the mass inflow rate only reaches to about $10^{-5}~M_\odot$ yr$^{-1}$ at the end of this run. The mass inflow rate in the $\beta$ model when the cooling flow reaches the quasi-steady state is also very low, $\sim 0.02 M_\odot$ yr$^{-1}$, which is due to the fact that the total gas mass in the Galactic halo in this model is small (less than $10\%$ of $M_{\rm mbar}$) and is consistent with the estimate in \citealt{MillerBregman2015}. The mass inflow rate would increase as the total halo gas mass in the $\beta$ model increases, similar to the result shown in Section 3.2. For the NFW and cored-NFW models, central gas densities are quite high, which explains why the cooling catastrophe starts very early in these runs. Since the total halo gas masses are quite high ($M_{\rm g}= M_{\rm mbar}$) in these two runs, the maximum mass inflow rates across the inner boundary are also quite large ($\sim 70-90 M_\odot$ yr$^{-1}$). The decrease in the mass inflow rate at later times is caused by the slowly depletion of the gas in our computational domain due to inflows across the inner boundary. Note that the gas is not allowed to freely flow into our computational domain at the outer boundary. 

\section{summary and discussions}
\label{section:summary}

Theoretical and observational arguments suggest that a hot, extended gaseous halo exists around the MW. Observations imply that the gas density is around $10^{-5} - 10^{-3}$ cm$^{-3}$, and the plasma temperature is $\sim 10^6$ K along individual lines of sight. However, the spatial distribution of the hot gas, and its total mass are not well constrained so far.
The understanding of the density distribution of the hot gaseous halo can help to solve the missing baryon problem of the MW. In this work, we propose a new general model for the gas density distribution in the MW's halo, which typically lies between the centrally-peaked NFW model and the very extended MB model. By properly choosing the model parameters, our model can also approach the cored-NFW model and the MB model.

We investigate the evolution of the hot gas in the MW's halo under radiative cooling with a series of 2D hydrodynamic simulations started from hydrostatic equilibrium. We did a parameter study in our simulations, investigating the roles of our model parameters and gas metallicity. Our simulations clearly indicate that the total gas mass within the halo and gas metallicity play crucial roles on the mass inflow rate in the developed cooling flow, which increases with both the halo gas mass $M_{\rm g}$ and metallicity $Z$. For a typical metallicity $Z=0.3Z_\odot$, the mass inflow rate across the inner boundary of $1$ kpc increases from $\sim 5~M_{\odot}$ yr$^{-1}$ when $M_{\rm g}=0.3M_{\rm mbar}$ to $\sim 50~M_{\odot}$ yr$^{-1}$ when $M_{\rm g}=M_{\rm mbar}$, much larger than the SFR observed in the MW. This suggests that stellar and/or AGN feedback processes may play important roles in the evolution of the MW by heating the halo gas and suppressing cooling flows.

For a fixed total gas mass in the halo, the spatial distribution of the halo gas does not substantially affect the mass inflow rate after the cooling flow reaches the quasi-steady state, but it does significantly affect the onset time of the central cooling catastrophe. When the halo gas distribution becomes more centrally-peaked (e.g., for smaller values of $r_{1}$ and $r_{2}$ or larger values of $\alpha_{1}$ in Equation \ref{ourmodel}), the central gas cooling time becomes shorter and the central cooling catastrophe starts earlier. But the mass inflow rate in the developed cooling flow does not change much if the total halo gas mass is fixed. We also investigate the impact of the gravity from the Galactic disk and bulge on the evolution of the halo gas. For the same gas density distribution, the gravity from the disk and bulge increases the equilibrium gas temperatures in inner regions and thus delays the onset of the central cooling catastrophe, but it does not substantially affect the final mass inflow rate in the cooling flow.

We also investigate the development of the Galactic cooling flow with four other gas density models adopted from the literature: the MB model, the $\beta$ model, the NFW model, and the cored-NFW model, and confirm our results. In the MB model, the gas distribution is most spatially extended, and the central gas cooling time is even longer than our simulation time of about $1.5$ Gyr, at which the cooling catastrophe has not yet started. In the $\beta$ model directly adopted from \citet{MillerBregman2015}, the halo gas mass is quite low, leading to a small mass inflow rate ($\dot{M} \sim 0.02 M_\odot$ yr$^{-1}$) at the quasi-steady state. In the NFW or the cored-NFW model, the gas distribution is centrally peaked, resulting in a very short onset time of the central cooling catastrophe and the final mass inflow rate is quite large ($\gtrsim 50M_\odot$ yr$^{-1}$). Future X-ray observations with higher sensitivity and spectral resolution, particularly of nearby MW-type galaxies, can help better constrain the spatial distribution of the hot circumgalactic medium, and the importance of cooling flows and feedback processes on the evolution of MW-type galaxies. 

The importance of radiative cooling in the hot gaseous halo may also be characterized by the radiative power within the virial radius $r_{\rm vir}$, defined as
\begin{equation}
 P_{\rm rad}=\int_{r_{\rm min}}^{r_{\rm vir}} \int_{0}^{\pi} 2 \pi r^2 \mathcal{C} \sin \theta dr d\theta {.}
\end{equation}
At $T \sim 10^6$ K, the hot gas mainly emits in ultraviolet (UV) and soft X-rays, and the UV emission dominates. Obviously, the radiative power $P_{\rm rad}$ increases with the halo gas mass $M_{\rm g}$ and the gas metallicity $Z$. For given values of $M_{\rm g}$ and $Z$, $P_{\rm rad}$ increases gradually as the gas density distribution becomes more centrally peaked.  For the halo gas model adopted in our simulations with a typical metallicity $Z=0.3Z_\odot$, $P_{\rm rad}$ increases from $1.05\times 10^{40}$ erg/s when $M_{\rm g}=0.1M_{\rm mbar}$ to $9.46\times 10^{40}$ erg/s when $M_{\rm g}=0.3M_{\rm mbar}$ to $2.63\times 10^{41}$ erg/s when $M_{\rm g}=0.5M_{\rm mbar}$ to $1.06\times 10^{42}$ erg/s when $M_{\rm g}=M_{\rm mbar}$. Note that $P_{rad} \propto M_{\rm g}^2$ for a given $Z$.

The hot halo gas is expected to be heated by stellar and AGN feedback processes. Assuming the Galactic supernova rate of 1.9 events per century \citep{Diehl2006}, and a characteristic energy output of $10^{51}$ erg per supernova (\citealt{Ciotti1991}; \citealt{LiTonnesen2019}), the average heating rate from supernova feedback is about $6.03\times 10^{41}$ erg/s, which is more than enough to offset radiative cooling in most models unless $M_{\rm g} \gtrsim 0.76M_{\rm mbar}$. The heating rate from AGN feedback in the Galaxy is harder to constrain from current observations. Assuming that AGN feedback events similar to the Fermi bubbles \citep{su2010} happen in the Galaxy every 50 Myr and the energy output from each event is around $2 \times 10^{55}$ erg (\citealt{zhang20}; \citealt{guo12}), the average AGN feedback heating rate is $1.27\times 10^{40}$ erg/s. Although this is much less than the average supernova feedback heating rate, AGN feedback deposits energy to much larger regions in the halo, potentially having significant effects on the transport and mixing of the gas and metals in the halo.

We thank an anonymous referee for very helpful comments. X.-E. F. thanks Xin-Yue Shi for very useful discussions and Zhen-Yi Cai for the help on scientific plotting. This work was supported by National Natural Science Foundation of China 
(No. 11873072, 11633006, 11725312, 11421303), Natural Science Foundation of Shanghai (No. 18ZR1447100), and Chinese Academy of Sciences through the Key Research Program of Frontier Sciences (No. QYZDB-SSW-SYS033 and QYZDJ-SSW-SYS008). The numerical calculations in this paper have been done on the supercomputing system in the Supercomputing Center of University of Science and Technology of China.


\bibliography{ms.bib}

\end{document}